\documentclass[aps,floatfix,showpacs,nofootinbib,twocolumn]{revtex4-1}
\usepackage{amsmath}
\usepackage{graphics,epsfig}
\usepackage{graphicx}
\usepackage{longtable}

\begin{document}

\title{Relativistic Aberration and the Cosmological Constant in Gravitational Lensing\\ I: Introduction}

\author{Dmitri Lebedev}
\email{dmitri.lebedev@queensu.ca}

\author{Kayll Lake}
\email{lakek@queensu.ca}

\affiliation{Department of Physics, Queen's University, Kingston,
Ontario, Canada, K7L 3N6 }

\date{\today}

\begin{abstract}
An analysis of null geodesics in Schwarzschild de Sitter space is presented with special attention to their global `bending angles', local measurable angles, and the involvement of the cosmological constant. We make use of a general technique which allows for finding observable intersection angles of null trajectories analytically. A general relativistic aberration relationship is established as one of its applications. The question of whether or not the cosmological constant, $\Lambda$, contributes to orbits of light and to related observable quantities is addressed in detail. We also discuss the ongoing debate on this issue and respond to some recent papers on the topic. The dependence of measurable quantities on the motion of observers is stressed throughout. Exact formulas for measurable intersection angles, as well as gravitational lens equations for observers in the Schwarzschild de Sitter background are provided.
\end{abstract}

\maketitle

\section{Introduction}
Recently there has been much activity in re-examining the role of the cosmological constant, $\Lambda$, in gravitational lensing. In 1983 Islam, \cite{islam}, suggested that the trajectory of light in Schwarzschild-de Sitter, henceforth SdS, space is independent of $\Lambda$. Following his publication, until recently it was generally believed that $\Lambda$ does not contribute to gravitational deflection of light. In 2007 Rindler and Ishak, \cite{ri1}, considered the role of local angles in a simple setup of gravitational deflection of light in SdS space. They argued that although $\Lambda$ may not contribute to the orbit of a light ray, it does contribute to its total deflection angle. Rindler and Ishak's conclusions immediately led to both enthusiasm and scepticism. Among the many responses to their original paper, some authors searched for other ways in which the contribution of $\Lambda$ emerges, in support of Rindler and Ishak's conclusions, see for example \cite{sereno1,schucker1,kcd,bbs,bp,hammad,abb,zt}. Others tried to find errors in Rindler and Ishak's work and explain the invalidity of their conclusions, \cite{kp,park,simpson,ak,piattella,butcher}. Up until now there seems to be no clear consensus as to whether or not $\Lambda$ indeed plays a role in gravitational lensing. However, the papers that followed \cite{ri1} amount to an interesting investigation of the subject, to which we intend to contribute.

Inspired by Rindler and Ishak's work, we take a closer look at local (observable) angles. Specifically, the influence of $\Lambda$ on observable angles between intersecting null geodesics in SdS space. Complementary to Rindler and Ishak's investigation, we incorporate the general motion of observers into the analysis, and explain in full detail that while $\Lambda$ may have no direct affect on orbits of light, it can still affect measurements made on them. We point out some subtle issues that may be a source of disagreement in recent literature. Along the course of our presentation we employ a general formula for expressing observable intersection angles. This formula is essential in the topic at hand, but quite surprisingly is missing from recent literature. We also involve  the effects of relativistic aberration of light in our discussion, and demonstrate that the general angle formula encompasses this phenomenon. In fact, the formula can be used to derive an invariant aberration equation, applicable to any background geometry and orientation. This general equation reduces to the well-known aberration equation as a special case.

Employing Kottler coordinates, we begin by fully examining the dependence of null geodesics and light ray orbits on $\Lambda$. We discuss the role that initial conditions play and how it may influence the conclusions made. We then digress to derive some general results to help us analyze local angles in the current context of light rays in SdS space. Throughout our presentation of the topic, we try to shed light on possible areas of confusion and ambiguity, especially in the use of parameters such as the ``impact parameter" and the ``bending angle". We present some expressions of measurable angles in SdS space for some specific observers. Finally, putting together the concepts of local and global angles we derive the single source gravitational lens equation in SdS background. We compare our work to numerous results from recent publications and discuss the state of the topic so far. See \cite{ll} for detailed discussions on some of the issues addressed here.

\section{Light rays in spherical symmetry} \label{s2}
In this paper, we study the role of $\Lambda$ in the simplest case of lensing phenomenon. It will serve as a demonstration of the effect of $\Lambda$, and as a starting point for further investigation. Consider the spherically symmetric and static metric, with a corresponding line element of the form
\begin{equation}
ds^2 = -f(r)dt^2+\frac{dr^2}{f(r)}+r^2\sin^2(\theta)d\phi^2+r^2 d\theta^2, \label{linel}
\end{equation}
in the region where $f(r)>0$. In this manifold, the Euler-Lagrange equations give the following system for the geodesics:
\begin{align}
&\frac{d}{d\lambda}\left(-f(r)\dot{t}\right)=0,\\
&\frac{d}{d\lambda}\left(r^2 sin^2(\theta)\dot{\phi}\right)=0,\\
&\frac{d}{d\lambda}\left(r^2\dot{\theta}\right)=r^2 sin(\theta)cos(\theta)\dot{\phi}^2,\\
&\frac{d}{d\lambda}\left(\frac{\dot{r}}{f}\right)=-\frac{f'\dot{t}^2}{2}-\frac{f' \dot{r}^2}{2f^2}+rsin^2(\theta)\dot{\phi}^2+r\dot{\theta}^2,
\end{align}
where the dot represents differentiation with respect to the affine parameter $\lambda$, and the prime represents differentiation with respect to $r$. Consider a light ray with initial conditions $(t_0,r_0,\phi_0,\theta_0)$ and $(\dot{t}_0,\dot{r}_0,\dot{\phi}_0,\dot{\theta}_0)$. It is important to note that not all of the parameters $\dot{t}_0$, $\dot{r}_0$, $\dot{\phi}_0$ and $\dot{\theta}_0$ can be considered independent. The null geodesic is subject to the condition
\begin{equation}
0=-f(r)\dot{t}^2+\frac{\dot{r}^2}{f(r)}+r^2\sin^2(\theta)\dot{\phi}^2+r^2 \dot{\theta}^2, \label{nulcond}
\end{equation}
which means that only three of the $\dot{t}_0$, $\dot{r}_0$, $\dot{\phi}_0$ and $\dot{\theta}_0$ can be freely set. Let us concentrate on the case where $\dot{r}_0$, $\dot{\phi}_0$ and $\dot{\theta}_0$ are the free parameters, and the given set of initial conditions is $(t_0,r_0,\phi_0,\theta_0 ; \dot{r}_0,\dot{\phi}_0,\dot{\theta}_0)$. These initial conditions set an event in the manifold and a spacelike direction at the event. We will refer to them as `standard' initial conditions. Furthermore, due to spherical symmetry, without loss of generality we can set $\theta_0=\frac{\pi}{2}$ and $\dot{\theta}_0=0$. For such initial conditions we immediately find that $\ddot{\theta}=0$, which means that $\dot{\theta}=0$ and $\theta=\frac{\pi}{2}$. Therefore, the ray of light is confined to the subspace $\theta=\frac{\pi}{2}$ and the system then reduces to:
\begin{align}
&f(r)\dot{t}=f(r_0)\dot{t}_0=\sqrt{\dot{r}_0^2+r_0^2 f(r_0) \dot{\phi}_0^2} \label{tdot}, \\
&r^2 \dot{\phi}=r_0^2 \dot{\phi}_0 \label{phidot}, \\
&\ddot{r}=\frac{r_0^4 \dot{\phi}_0^2}{r^2}\left(\frac{f}{r}-\frac{f'}{2}\right) \label{rddot}.
\end{align}
From \eqref{nulcond}, \eqref{tdot} and \eqref{phidot} we obtain the first integral for $r$:
\begin{equation}
\dot{r}^2=\dot{r}_0^2 + r_0^4\dot{\phi}_0^2\left(\frac{f(r_0)}{r_0^2}-\frac{f(r)}{r^2}\right). \label{rdot}
\end{equation}
Our goal is to analyze the influence of the cosmological constant, $\Lambda$, on light rays. We set attention to Schwarzschild-de Sitter spacetime and make use of the Kottler metric, \cite{kot}. The metric has the form \eqref{linel} with
\begin{equation}
f(r)=1-\frac{2m}{r}-\frac{\Lambda}{3} r^2, \label{fsds}
\end{equation}
where $m$ is a mass parameter, the effective gravitational mass being $M=m+\frac{\Lambda}{6}r^3$, \cite{lake}. With this $f(r)$, we find that $\Lambda$ gets cancelled out in the differential equations for $r$, leaving $m$ as the only metric parameter present. From equation \eqref{tdot} it follows that the evolution of $t$ depends on both $m$ and $\Lambda$, while from \eqref{phidot} and \eqref{rdot} with \eqref{fsds}, the evolutions of $r$ and $\phi$ depend only on $m$ and the initial conditions. We conclude as follows: In Schwarzschild-de Sitter spacetime, described by the Kottler metric, the evolutions of the spatial coordinates of a light ray with standard initial conditions do not depend on $\Lambda$. It should be clear that it is a coordinate dependent result, true in Kottler coordinates, for rays with standard initial conditions. We can obtain a first order differential equation in $r$ and $\phi$ by using equations \eqref{phidot} and \eqref{rdot},
\begin{equation}
\left(\frac{dr}{d\phi}\right)^2=\frac{r^4 \dot{r}_0^2}{r_0^4 \dot{\phi}_0^2}+r^4\left(\frac{f(r_0)}{r_0^2}-\frac{f(r)}{r^2}\right). \label{orbfir}
\end{equation}
This equation and its solution (the spatial orbit) do not include $\Lambda$ explicitly. Therefore, we completely agree with Islam's remarks on the absence of $\Lambda$ in \cite{islam}, and with Rindler and Ishak's position on the issue in \cite{ri1}. However, it must be pointed out that Islam merely reports of the absence of $\Lambda$ from the second order differential equation in $u=\frac{1}{r}$ and $\phi$. This absence in the second order relationship is simply not enough to firmly conclude the non-influence of $\Lambda$ on the orbit. Indeed, we can differentiate \eqref{orbfir} in such a way that will eliminate $m$, leaving it absent from the resulting second order equation \footnote{It is common to convert equation \eqref{orbfir} (with \eqref{brel}) to the familiar $\frac{d^2 u}{d\phi^2}=3mu^2-u$, by setting $u=\frac{1}{r}$ and differentiating. Notice that by setting $u=\sqrt{r}$ instead, we get $\frac{d^2 u}{d\phi^2}=\frac{3u^5}{4}\left( \frac{1}{b^2}+\frac{\Lambda}{3} \right)-\frac{u}{4}$.}. But this doesn't mean that $m$ has no influence on the orbit. It will make its way back to the solution once integration is done and initial (or boundary) conditions are applied. Thus, the claim that $\Lambda$ has no influence on orbits of light entirely based on Islam's findings is justly subject to scepticism (see for example \cite{sereno1,bbs,ak,bp}). The invalidity of Islam's argument was stated explicitly in \cite{lake}. To make this issue finally clear, we arrive at the above result from first principles and explicitly state under what conditions it holds. It is worth noting that for $f(r)$ given by \eqref{fsds}, in the combination $\left( \frac{f(r_0)}{r_0^2}-\frac{f(r)}{r^2} \right)$ we have a complete cancellation of $\Lambda$. Using approximations that tamper with this combination may affect the cancellation, \cite{sereno1}.

In the literature on the subject it is common to make use of the parameter $b$ instead of the initial conditions $r_0$, $\dot{r}_0$ and $\dot{\phi}_0$ in \eqref{orbfir}. $b$ is defined as the ratio between the conserved angular momentum and energy of the photon. In terms of our parameters it can be written as
\begin{equation}
b=\frac{r_0^2 \dot{\phi}_0}{f(r_0)\dot{t}_0}, \label{bdef}
\end{equation}
or with \eqref{tdot}
\begin{equation}
\frac{1}{b^2}=\frac{\dot{r}_0^2}{r_0^4 \dot{\phi}_0^2}+\frac{f(r_0)}{r_0^2}. \label{brel}
\end{equation}
Treating $b$ as an independent parameter in the resulting differential equation leads to the explicit dependence of orbits on $\Lambda$. It is equivalent to treating $\dot{t}_0$ and $\dot{\phi}_0$ instead of $\dot{r}_0$ and $\dot{\phi}_0$ as the independent parameters, as is clear from \eqref{bdef}. We see that putting $b$ in \eqref{orbfir} leads to the absorption of $\frac{f(r_0)}{r_0^2}$, leaving the $\Lambda$ present in $\frac{f(r)}{r^2}$ without cancellation. We note again the importance of specifying the independent parameters of the orbit before making conclusions regarding the influence of $\Lambda$ on it.

Another commonly used parameter is the `impact parameter' $B$. It is related to $b$ through
\begin{equation}
\frac{1}{B^2}=\frac{1}{b^2}+\frac{\Lambda}{3}, \label{Bdef}
\end{equation}
and if used in \eqref{orbfir} it completely absorbs the explicit appearance of $\Lambda$. It turns out that $\Lambda$ is not involved in the relationship between $B$ and the standard initial conditions either. If $B$ is treated as an independent parameter in the differential equation, the resulting orbits do not depend on $\Lambda$ explicitly. Again, the dependence of the orbits on $\Lambda$ is entirely associated with the choice of initial conditions or the parameters used. We have summarised some sensitive aspects regarding the effect of $\Lambda$ on light rays in spherical symmetry which were not discussed in recent papers on the topic. Some of the disagreements in papers that followed Rindler and Ishak's \cite{ri1} are related to these aspects. We present a much more detailed discussion on this in \cite{ll}.

It is worth pointing out that the two traditional definitions of the `impact parameter' in Schwarzschild space do not coincide in SdS space \footnote{The impact parameter of an orbit in central potential is roughly defined as either the ratio between the conserved angular momentum and energy, or as the distance of an asymptote to the origin.}, see \cite{hartle,wald,weinberg,mtw,ri1,kkl}. This term has been used to refer to both $b$ and $B$ by different authors (e.g. \cite{sereno1,bbs,hammad}, \cite{ri1,kp,ak}), which are clearly different quantities for $\Lambda \not= 0$. We adhere to the geometrical definition, in which the impact parameter is the geometrical distance of an asymptote of the orbit to the origin. This is found to be $B$ with $\Lambda \not= 0$, not $b$, see \cite{ll}.

Rindler and Ishak, in agreement with Islam's claim, have acknowledged that $\Lambda$ has no influence on the orbit itself. However, they argued that since $\Lambda$ affects the global geometry of the space in which the orbit exists, it must have an effect on the orbit's overall bending. They proposed and demonstrated the use of a new definition for the bending angle of the orbit, which is affected by $\Lambda$. More on this in section \ref{s4a}. Other attempts in extending the concept of bending angle to the $\Lambda \not= 0$ case were made since. There is still disagreement between authors on the exact definition.

Here we adopt a different approach and make a clear distinction between global and local angles. We review the purpose of the bending angle defined in the absence of $\Lambda$, before attempting its generalization for $\Lambda \not= 0$. Keeping in mind that the basic question of theoretical interest here is whether or not $\Lambda$ affects directly observable angles, and if so, can its value be determined from such measurements? In addressing these questions, we proceed with a method that allows us to account for the motion of observers making local measurements and, in particular, to study the combined effects of $\Lambda$ and relativistic aberration on observable angles.

\section{Intersection angles and Relativistic aberration of light} \label{s3}
Consider a spacetime with a metric $g_{\alpha \beta}$ of signature $+2$, and two null geodesics $\Gamma^1$ and $\Gamma^2$ that intersect at an event $p$. In general, for an indefinite metric the angle between two arbitrary vectors is not defined, nor is the intersection angle between our $\Gamma^1$ and $\Gamma^2$. However, if we consider a timelike observer at the intersection event $p$, then the measurable intersection angle that takes place in the local frame of the observer has a clear definition. Let $K$ and $W$ be 4-vectors at $p$ tangent to $\Gamma^1$ and $\Gamma^2$ respectively, and $U$ be the 4-velocity of an observer at $p$. The intersection angle, $\psi_U$, between the light rays measured by the observer is given by
\begin{equation}
\cos(\psi_U)=\frac{g_{\alpha \beta}K^\alpha W^\beta}{(g_{\alpha \beta}U^\alpha K^\beta)(g_{\alpha \beta}U^\alpha W^\beta)}+1. \label{form}
\end{equation}
The derivation of this formula is trivial and its importance to the field of gravitational lensing is clear, yet it is not utilized in any of the recent literature on the topic. The formula is known to be used by the GAIA team. We found an explicit appearance of it in \cite{firstform} and an implicit form in \cite{defe1,defe2}. We were not able to find the formula in books on differential geometry, except for very special cases appearing in \cite{oneill,kriele}. Also see \cite{ellis,cohen,ehlersform}.

A simple derivation follows. For a spacetime with a metric $g_{\alpha \beta}$ of signature $+2$, let $K$ and $W$ be any two 4-vectors at  an event $p$, each may represent the tangent of an arbitrary curve through $p$. Let $U$ be the 4-velocity of a timelike observer at $p$, and $\overline{K}$ and $\overline{W}$ be the projections of $K$ and $W$ onto the local space of the observer, such that $\overline{K}$ and $\overline{W}$ are spacelike and perpendicular to $U$ with respect to the metric $g_{\alpha \beta}$. We easily find that
\begin{align}
&\overline{K}^\alpha=K^\alpha+K^\beta U_\beta U^\alpha\\
&\overline{W}^\alpha=W^\alpha+W^\beta U_\beta U^\alpha.
\end{align}
Projecting the metric $g_{\alpha \beta}$ onto the spatial frame of the observer at $p$ will give the positive definite (local) metric of that space, which must be equivalent to the confinement of $g_{\alpha \beta}$ to 4-vectors tangent to that spatial frame. The angle, $\psi_U$, between the vectors $\overline{K}$ and $\overline{W}$ that takes place in the observers space is clearly given by
\begin{equation}
\cos(\psi_U)=\frac{g_{\alpha \beta} \overline{K}^\alpha \overline{W}^\beta}{\sqrt{g_{\alpha \beta} \overline{K}^\alpha \overline{K}^\beta}\sqrt{g_{\alpha \beta} \overline{W}^\alpha \overline{W}^\beta}}.
\end{equation}
Using the expressions for $\overline{K}$ and $\overline{W}$ we get
\begin{equation}
\cos(\psi_U)=\frac{K_\alpha W^\alpha + (U_\alpha K^\alpha)(U_\beta W^\beta)}{\sqrt{K_\alpha K^\alpha + (U_\alpha K^\alpha)^2} \sqrt{W_\alpha W^\alpha + (U_\alpha W^\alpha)^2}}.
\end{equation}
This formula gives the measurable intersection angle between any curves with corresponding tangent 4-vectors $K$ and $W$. In the special case where the curves are null geodesics, the formula simplifies to \eqref{form}. For more details see \cite{ll}.

Turning back attention to light rays in SdS space, we see that although $\Lambda$ may not affect the orbit in the $(r,\phi)$ coordinate plane, the local (observable) angles have a clear dependence on both the metric $g_{\alpha \beta}$ and the motion of the observer $U^\alpha$, either of which may carry $\Lambda$ explicitly.

Before ending this section, we consider the connection between formula \eqref{form} and relativistic aberration of light. As it is well known from special relativity, and clearly suggested by \eqref{form}, two observers in relative motion will generally measure different angles between two intersecting light rays at an event. For a particular orientation of the motions of the rays and observers, the relationship between the two measurable angles, $\psi_1$ and $\psi_2$, is given by the relativistic aberration equation:
\begin{equation}
\cos(\psi_2)=\frac{\cos(\psi_1)-v}{1-v\cos(\psi_1)}, \label{specrelab}
\end{equation}
where $v$ is the relative speed between the observers. The derivation of this formula can be found in most texts on special relativity. The assumed orientation is such that in the rest frame of one observer the direction of travel of the other coincides with the direction of travel of a light ray. It is advantageous to have a fully general relationship between the measurable angles for any orientation in terms of the four 4-vectors involved. We can easily obtain such a relationship from \eqref{form}. If $K$ and $W$ are 4-vectors tangent to the intersecting light rays, and $U$ and $V$ are the 4-velocities of two observers that measure the intersection angles $\psi_U$ and $\psi_V$ respectively, then we have
\begin{equation}
\frac{\cos(\psi_U)-1}{\cos(\psi_V)-1}=\frac{(g_{\alpha \beta}V^\alpha K^\beta)(g_{\alpha \beta}V^\alpha W^\beta)}{(g_{\alpha \beta}U^\alpha K^\beta)(g_{\alpha \beta}U^\alpha W^\beta)}. \label{genrelab}
\end{equation}
The above is a fully general formula for relativistic aberration that can be used with any metric and any orientation of motion. To see that \eqref{specrelab} is a special case of \eqref{genrelab} we can restrict ourselves to the situation where the directions of travel of one the observers and one of the light rays coincide in the local frame of the other observer. In a more technical language, this special case is summarized by the condition that the 4-vectors $U$, $V$ and one of the light rays, say $K$, are not linearly independent. Explicitly,
\begin{equation}
K^\alpha = c_1 U^\alpha + c_2 V^\alpha, \label{cond1}
\end{equation}
where the constants $c_1$ and $c_2$ are determined from the inner products of the vectors involved. Under the conditions $U^\alpha U_\alpha = V^\alpha V_\alpha = -1$, $K^\alpha K_\alpha = 0$, and $U^\alpha V_\alpha = -\gamma$, where $\gamma$ is the special relativistic factor, we find the ratio
\begin{equation}
\frac{c_1}{c_2} = \gamma v - \gamma. \label{cond2}
\end{equation}
Evaluating the inner products in \eqref{genrelab} by using \eqref{cond1} and \eqref{cond2} yields \eqref{specrelab}, as required. This is a coordinate free derivation of the aberration equation.

Equations \eqref{form} and \eqref{genrelab} have a particularly simple form in the small angle limit. Notice the effect of the signature of the metric in the derivation and outcome of equation \eqref{form}, but not in \eqref{genrelab}.

\section{Bending of light and gravitational lensing in SdS} \label{s4}
In this section we will begin by clearing up some ambiguity regarding the bending angle parameter and settle on a suitable definition of it. In doing so, we will consider that the main purpose of this angle in SdS will be the same as it is in Schwarzschild space, which is to be used in the derivation of the single source gravitational lens equation. We will proceed by collecting some expressions for local angles and combine the results to produce a single source lens equation in a SdS background.

\subsection{Defining the bending angle with $\Lambda \not= 0$} \label{s4a}
\begin{figure}[!ht]
\epsfig{file=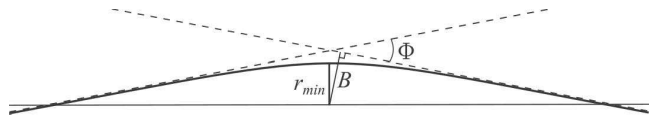,width=85mm}
\caption{Symmetric bending orbit of light in either Schwarzschild or SdS space on the ($r,\phi$) coordinate plane. With smallest $r$ ($r_{min}$), impact parameter ($B$), and geometric bending angle ($\Phi$).} \label{fig1}
\end{figure}

We start by reviewing the concept of `bending angle' in asymptotically flat Schwarzschild space and proceed to discuss its generalization to SdS space. The bending angle for a symmetric orbit of a light ray in Schwarzschild space is defined and calculated in various ways throughout the literature, \cite{rindler,weinberg,ehlers,mtw,peacock,hendriksen,hartle,kriele,dinverno,stephani,wald,pk,irwin}. Within the different definitions, the expressions for the bending angle all coincide. With reference to Figure \ref{fig1}, the bending angle is most commonly defined as the angle between the asymptotes of the orbit, $\Phi$, \cite{kriele,dinverno,stephani,wald,pk,irwin}. For an orbit with smallest $r$ coordinate $r_{min}$ and impact parameter $b$ (=$B$ for $\Lambda=0$), the bending angle in Schwarzschild space is found to be
\begin{equation}
\Phi=\frac{4m}{r_{min}}+ \mathcal{O} \left(\frac{m}{r_{min}}\right)^2=\frac{4m}{b}+ \mathcal{O} \left(\frac{m}{b}\right)^2.
\end{equation}

A natural method to find this expression is to take the limits of $r \rightarrow \infty$ in the orbit solution, a method which comes under scrutiny when extending the concept to SdS space, \cite{ri1,bbs,bp}. Although $\Lambda$ may not affect the orbit itself, with its presence in the metric there is no asymptotic flatness, and the region of interest is bounded by an outer horizon.

In the pioneering attempt to modify the definition of the bending angle for the $\Lambda \not= 0$ case in \cite{ri1}, the authors pointed out the problem with taking the limit and proposed a definition that involves a local angle. By using the appropriate metric to express this local angle, they found that $\Lambda$ appears explicitly in the expression of the bending angle, see equation (17) in \cite{ri1}. This modified expression reduces to the known bending angle in Schwarzschild for $\Lambda=0$ and can be viewed as a generalization. In this new definition, the observers that could in principle measure the local angle are static, although this is not explicitly stated by the authors. This definition has been criticized for different reasons, \cite{kp,park,simpson,bbs,ak,bp,butcher}. See Ishak's et al. follow-up papers on the topic, \cite{ri2,ridma,ishak1,rid}, and further discussion in \cite{ll}. Other authors have proposed adjustments to the bending angle that also result in the explicit appearance of $\Lambda$ in its expression, e.g. \cite{sereno1,schucker1,bbs,bp,ak}, most are modifications of the definition in \cite{ri1}.

\begin{figure}[!ht]
\epsfig{file=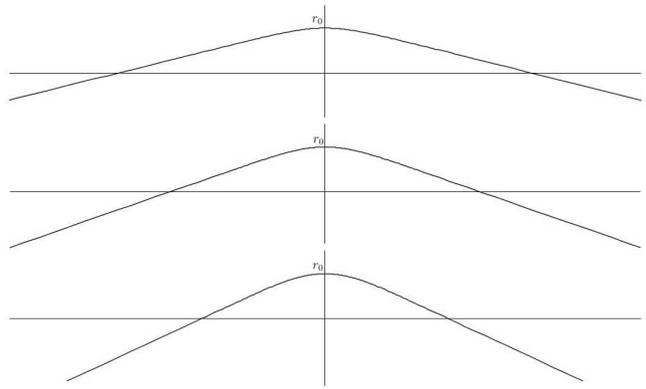,width=85mm}
\caption{Symmetric orbit of light in polar coordinates, with $r_{min}=r_0$. The value of $m$ is successively increasing top to bottom, $\Lambda$ is kept constant.} \label{fig2}
\end{figure}

\begin{figure}[!ht]
\epsfig{file=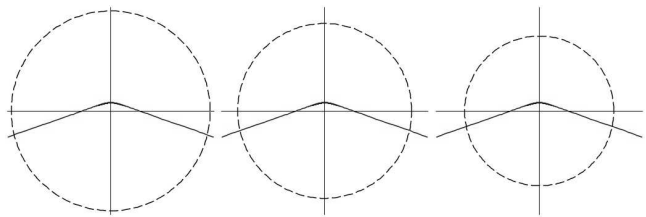,width=85mm}
\caption{Symmetric orbit of light in polar coordinates, varying $\Lambda$ and constant $m$. The outer horizon is shown, as it decreases with increasing $\Lambda$, the orbit remains unchanged.} \label{fig3}
\end{figure}

Consensus should be reached on an exact definition of the bending angle in SdS background. We recall that a key purpose of the bending angle is in deriving the gravitational lens equation, which is of most practical use. This is the equation where the explicit appearance of $\Lambda$ is of most significance. While the orbital equation, \eqref{orbfir}, holds no information about the outer horizon, it properly describe orbits within the horizon on the $(r,\phi)$ coordinate plane. The geometric angle between the asymptotes of a light orbit is the required quantity in the ray trace derivation of the lens equation. For this reason, we propose that the bending angle will be defined as the geometric angle between the asymptotes of an orbit, in both Schwarzschild and SdS backgrounds. The bending angle need not be directly measurable, it is a parameter of the orbit. In particular, with this definition, a symmetric orbit of light on the $(r,\phi)$ coordinate plane, will have three related parameters: the closest approach $r_{min}$, the impact parameter $B$ (or $b$ for $\Lambda=0$), and the bending angle $\Phi$. In SdS space, the angle between the asymptotes of an orbit of light is
\begin{equation}
\Phi=\frac{4m}{r_{min}}+ \mathcal{O} \left(\frac{m}{r_{min}}\right)^2=\frac{4m}{B}+ \mathcal{O} \left(\frac{m}{B}\right)^2.
\end{equation}
In summary, the influence of varying $m$ and $\Lambda$ in the situation of interest is depicted in Figures \ref{fig2} and \ref{fig3}. The role of $\Lambda$ in gravitational lensing will be attributed to its influence on the local angles involved in the lens equation.

\subsection{Gravitational lens equation} \label{sec:lens}
\begin{figure}[!ht]
\epsfig{file=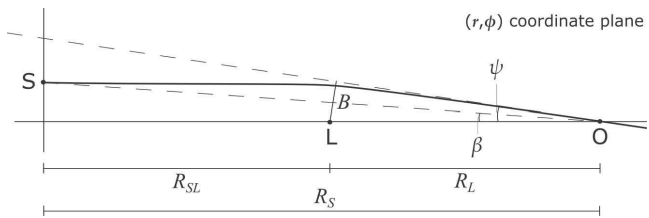,width=85mm }
\caption{Standard setup of gravitational lensing with  source (S), observer (O), and lens (L).} \label{lensing}
\end{figure}

We base our discussion of gravitational lensing and the assumptions involved mainly on \cite{ehlers}, but also see \cite{rindler,hartle,irwin,pk,peacock,stephani,bozza}. The standard setup of source, lens, and observer is depicted in Figure \ref{lensing}. In principle, one could use equation \eqref{orbfir} to get an exact relationship between the geometric angles $\psi$ and $\beta$. But if the source and observer are located in the asymptotic region of the orbit, and all the associated angles are small, then the `ray-trace' derivation can be used, \cite{ehlers,hartle,irwin}. In this procedure, with reference to the parameters appearing in Figure \ref{lensing}, we find
\begin{equation}
\psi R_S = \beta R_S + \Phi R_{SL},
\end{equation}
where $\Phi$ is explained in Figure \ref{fig1}.
For an orbit with large impact parameter $B$, to first order in $\frac{m}{B}$ and the angles, we get
\begin{align}
\beta &= \psi - \frac{4m R_{SL}}{B R_S}\\
&= \psi - \frac{4m R_{SL}}{\psi R_L R_S}. \label{lensg}
\end{align}

We now replace the geometric parameters with measurable ones by using angular diameter distances and equation \eqref{form}. Let $U$ be the 4-velocity of the observer making the measurements, $K$ and $W$ be the 4-vectors of the intersecting rays in equation \eqref{form}. Then, to first order in angles, the measurable angle $\psi_U$ is found to be
\begin{equation}
\psi_U = \frac{f(R_L)\psi}{f(R_L)U^t-U^r}, \label{lenspsi}
\end{equation}
with reference to the small geometric angle $\psi$ on the diagram. The non-radial velocity components of $U$ enter the above through $U^t$. Exact expressions for $\psi_U$ and without reference to the geometric $\psi$ are given in Appendix \ref{app:exact}. Let
\begin{equation}
h(r,U) = \left[ \frac{f(r)}{f(r)U^t-U^r} \right]_{m=0}. \label{lensh}
\end{equation}
In the absence of the lens, the position angle $\beta_U$, where the observer would have seen the source, is found to be
\begin{equation}
\beta_U = h(R_L,U) \beta \label{lensbeta}.
\end{equation}
In the cosmological context, the geometric `distances' $R_L$, $R_S$ and $R_{SL}$ should be replaced with available angular diameter distances. Let $D_L$ and $D_S$ be the measurable angular diameter distances to the lens and source, respectively. Adopting the usual definition, see for example Section 3.5 in \cite{ehlers}, to the current situation, we find
\begin{equation}
D_L = \frac{R_L}{h(R_L,U)}, \; \; D_S = \frac{R_S}{h(R_L,U)}, \label{lensd}
\end{equation}
and
\begin{equation}
R_{SL} = (D_S - D_L)h(R_L,U), \label{lensr}
\end{equation}
to first order in the angles. Using \eqref{lenspsi}, \eqref{lensbeta}, \eqref{lensd}, and \eqref{lensr} with \eqref{lensh} in \eqref{lensg}, we find to first order in the angles and mass
\begin{equation}
\beta_U = \psi_U - \frac{4m (D_S-D_L)}{D_S D_L \psi_U }h(R_L,U). \label{lensm}
\end{equation}
The above is the single source gravitational lens equation for the SdS background. We emphasise that this result can be derived without the use of bending angle at all, or any reference to geometric quantities in Figure \ref{lensing}; this alternative procedure is sketched in the Appendix \ref{app:lens}. Once an observer is set, the remaining $R_L$, which comes through $h(R_L,U)$ in \eqref{lensm}, must be replaced with $D_L$ to get the final result. This is done by solving \eqref{lensd}(i) for $R_L$. It should be noted that although approximations on angles and $m$ have been made, the above result is exact in $\Lambda$ and the components of $U$.

Two specific observers are of interest here. First, the static observer with $U^\alpha = \left( \frac{1}{\sqrt{f(R_L)}},0,0,0 \right)$, for which we get
\begin{equation}
\beta_U = \psi_U - \frac{4m (D_S-D_L)}{D_S D_L \psi_U \sqrt{1+\frac{\Lambda}{3} D_L^2}}.
\end{equation}
Next, in the cosmological setting, a comoving observer far away from the mass has 4-velocity $U^\alpha = \left( \frac{1}{1-\frac{\Lambda}{3}R_L^2}+ \mathcal{O} \left( \frac{m}{R_L}\right) ,\sqrt{\frac{\Lambda}{3}}R_L+\mathcal{O} \left( \frac{m}{R_L}\right),0,0 \right)$. In this case, with $\frac{\Lambda}{3}=H^2$, we find
\begin{equation}
\beta_U = \psi_U - \frac{4m (D_S-D_L)}{D_S D_L \psi_U (1-HD_L)}. \label{lenscom}
\end{equation}
The above result can be interpreted as the single source cosmological lens equation for SdS background, where all the parameters are measurable by a comoving observer. It is in agreement with the main result in \cite{park}, see equation (29) there, which was derived through a different approach. Interestingly, the authors of \cite{park} conclude that low orders of $\Lambda$ do not appear in the lens equation, but again we encounter a situation where the choice of parameters strongly affect such conclusions. The authors of \cite{park} make use of the angular diameter distance $d_{SL}$, found to be related to the distances $D_L$ and $D_S$ through (compare to (30) in \cite{park})
\begin{equation}
d_{SL} = \frac{D_S-D_L}{1-HD_L}.
\end{equation}
In principle $d_{SL}$ could be measured by a second (distant) observer, but in practice it may not be available directly. If all observations are assumed to be taken at a single point, as in the cosmological setting, then the angular diameter distance $d_{SL}$ must be determined indirectly, from other measurements. We find a complete cancelling of $\Lambda$ containing terms from equation \eqref{lenscom} when using the angular diameter distance $d_{SL}$. But if only $D_L$ and $D_S$ are directly available, then the value of $d_{SL}$ can be established only with knowledge of $\Lambda$. See \cite{ll} for further discussion of \cite{park}. We conclude that when expressing the cosmological lens equation in terms of directly measurable parameters, $\Lambda$ makes an explicit appearance.

An advantage in using Kottler coordinates is due to the fact that the problem is largely reduced to the orbital equation, \eqref{orbfir}. In this setting, the bending angle and other geometrical parameters come in handy. However, a disadvantage is apparent when trying to express angles measured by cosmological observers away from the origin; an obstacle easily overcame by utilizing formula \eqref{form}. The authors of \cite{park,kp,piattella} dealt with this problem by adopting to coordinates centred at the observer, introducing various complexities to the analysis. This illustrates the flexibility that the formula adds to the topic. Furthermore, using the Kottler metric allows us to closely follow the traditional analysis of lensing by a single source, as in \cite{ehlers}, and can be extended to produce a multi-source lens equation that models a more realistic situation. See Section 2.2 of \cite{ehlers}. Other approaches may present a much higher mathematical difficulty. We leave the derivation of multi-source lens equations for $\Lambda \not= 0$ background to future work.

\section{Conclusion}
It was a long time ago that the question of what exact role $\Lambda$ plays in gravitational lensing was asked. But even in the simplest special case of the problem, until this day, the topic seems to be suffering from ambiguities and disagreements. In the course of the ongoing investigation it became clear that in order to properly address the problem one must consider the contributions of the metric elements on local measurements, on top of the geodesic equations and the shape of a light orbit, \cite{ri1}.

Building up on previous work, in this paper we investigated the effects of $\Lambda$ as well as the motion of observers on local measurements. It was definitively concluded that $\Lambda$ does not appear in the orbital equation when considering `standard' initial conditions. We made use of the angle formula \eqref{form} to account for observer motion in this topic, demonstrated its connection with aberration, and employed it in the context of gravitational lensing. The formula allows more flexibility in the analysis and is a needed complement to the orbital equation when using Kottler coordinates. The complexities that arise in defining the bending angle and the impact parameter when $\Lambda \not= 0$ were discussed. Finally, by combining the concepts we showed that $\Lambda$ appears explicitly in the single source lens equation for SdS background. The importance of the choice of parameters used in the lens equation and the orbital equation were addressed as well.

In addition to the papers mentioned thus far, see \cite{mirag,lake2,sereno2,sereno3} and some references therein for more on this topic. A more detailed discussion of some of the most cited works on the topic, including \cite{ri1}, \cite{sereno1}, \cite{kp} and \cite{park} can be found in \cite{ll}. Beyond extending the analysis to multi-source lens equations, in follow up work we will address magnification effects, red-shift and luminosity distances, time delay effects, other contributing cosmological factors, and discuss how in practice $\Lambda$ can be measured through gravitational lensing phenomena \cite{2}.

\begin{acknowledgments}
This work was supported by a grant (to KL) from the Natural Sciences and Engineering Research Council of Canada.
\end{acknowledgments}

\appendix

\section{Exact expressions for measurable angles in SdS} \label{app:exact}
We express $\psi_U$ by means of \eqref{form}, with reference to the Kottler metric, where one of the 4-vectors $K$ or $W$, is radial and the other is subject to equations \eqref{tdot}, \eqref{phidot} and \eqref{rddot}. For coordinate $r$ of the observer, with $r_{min}=r_0$ for the orbit,

\begin{align}
&\cos(\psi_U) = \frac{g_{\alpha \beta}K^\alpha W^\beta}{(g_{\alpha \beta}U^\alpha K^\beta)(g_{\alpha \beta}U^\alpha W^\beta)}+1 = 1+\\ \label{formsds}
\nonumber &\frac{-\sqrt{\frac{f(r_0)}{r_0^2}}+\sqrt{\frac{f(r_0)}{r_0^2}-\frac{f(r)}{r^2}}}{(-\sqrt{\frac{f(r_0)}{r_0^2}} U^t+ \sqrt{\frac{f(r_0)}{r_0^2}-\frac{f(r)}{r^2}}\frac{U^r}{f(r)}+U^\phi) (-f(r)U^t+U^r)}.
\end{align}

In the above expression, $\Lambda$ comes in through $f(r)$ and $f(r_0)$. The components of $U$ are subject to $U^\alpha U_\alpha = -1$, $U^t$ can be replaced with spatial velocity components, introducing more metric terms. Other parameters related to the orbit may come in through $r_0$.

For static observer, with $U^\alpha = \left( \frac{1}{\sqrt{f(r)}},0,0,0 \right)$, we have
\begin{equation}
\cos(\psi_{static})=\frac{\sqrt{\frac{f(r_0)}{r_0^2}-\frac{f(r)}{r^2}}}{\sqrt{\frac{f(r_0)}{r_0^2}}},
\end{equation}
and
\begin{equation}
\tan(\psi_{static})=\frac{\sqrt{\frac{f(r)}{r^2}}}{\sqrt{\frac{f(r_0)}{r_0^2}-\frac{f(r)}{r^2}}}.
\end{equation}
Compare the above with equations (15,16) in \cite{ri1}.

For comoving observer far away from the mass, with $U^\alpha = \left( \frac{1}{1-H^2r^2},Hr,0,0 \right) = \left( \frac{1}{f_{m=0}(r)},Hr,0,0 \right)$, we find
\begin{equation}
\cos(\psi_{comoving}) = \frac{\sqrt{\frac{f(r_0)}{r_0^2}-\frac{f_{m=0}(r)}{r^2}}-\sqrt{\frac{f(r_0)}{r_0^2}}Hr}{\sqrt{\frac{f(r_0)}{r_0^2}}-\sqrt{\frac{f(r_0)}{r_0^2}-\frac{f_{m=0}(r)}{r^2}}Hr}.
\end{equation}
Notice how $\Lambda$ terms come in through the velocity components of the observer as well as the metric itself.

\section{Alternative derivation of the lens equation in SdS} \label{app:lens}
Through more laborious procedures, equation \eqref{lensm} can be derived without use of auxiliary parameters and the bending angle. We outlined an example below.

With reference to the definitions in Section \ref{sec:lens}, let the $(r,\phi)$ coordinates of observer and source be $(r_O,0)$ and $(r_S,\pi-\delta_S)$, respectively. The angle $\beta_U$ can be expressed by means of \eqref{form} and \eqref{orbfir} with $m=0$. To first order in $\beta_U$ and $\delta_S$, we find
\begin{equation}
\beta_U = h(r_O,U)\frac{r_S \delta_S}{r_O+r_S}, \label{applensbeta}
\end{equation}
where $h(r,U)$ is given by \eqref{lensh}. Next, we replace the coordinate parameters $r_O$, $r_S$, and $\delta_S$ with measurable position angle $\psi_U$, and angular diameter distances $D_L$ and $D_S$. To first order in angles and mass, by using equations \eqref{form} and \eqref{orbfir}, $\psi_U$ is expressed as
\begin{equation}
\psi_U = h(r_O,U)\left( \frac{r_S \delta_S}{r_O+r_S}+\frac{4m}{r_O \delta_S} \right). \label{applenspsi}
\end{equation}
The angular diameter distances can be found analytically through various methods, such as integrating the geodesic deviation equation. To first order in $\delta_S$ we get
\begin{equation}
D_L = \frac{r_O}{h(r_O,U)}, \; \; D_S = \frac{r_O+r_S}{h(r_O,U)}. \label{applensd}
\end{equation}
Using equations \eqref{applenspsi} and \eqref{applensd} in \eqref{applensbeta} yields \eqref{lensm}, as required. This procedure can be modified to yield higher order terms in the resulting relationship.


\begin{thebibliography}{}
\bibitem{islam} J. N. Islam, Phys.~Lett.~A \textbf{97}, 6, 239-241, (1983).
\bibitem{ri1} W. Rindler, M. Ishak, Phys.~Rev.~D \textbf{76}, 043006 (2007), [arXiv:0709.2948].
\bibitem{sereno1} M. Sereno, Phys.~Rev.~D \textbf{77}, 043004 (2008), [arXiv:0711.1802].
\bibitem{schucker1} T. Schucker, Gen.~Rel.~Grav. \textbf{41}, 7, 1595-1610, (2009), [arXiv:0807.0380].
\bibitem{kcd} R. Kantowski, B. Chen, X. Dai, Astrophys. J. \textbf{718}, 913 (2010), [arXiv:0909.3308].
\bibitem{bbs} A. Bhadra, S. Biswas, K. Sarkar, Phys.~Rev.~D \textbf{82}, 063003 (2010), [arXiv:1007.3715].
\bibitem{bp} T. Biressa, J. A. de Freitas Pacheco, Gen.~Rel.~Grav. \textbf{43}, 10, 2649-2659, (2011), [arXiv:1105.3907].
\bibitem{hammad} F. Hammad, Mod. Phys. Lett. A, \textbf{28}, 1350181 (2013), [arXiv:1309.0263].
\bibitem{abb} M. Aghili, B. Bolen, L. Bombelli, (2014), [arXiv:1408.0786].
\bibitem{zt} F. Zhao, J. Tang, Phys.~Rev.~D \textbf{92}, 083011 (2015).
\bibitem{kp} I. Khriplovich, A. Pomeransky, Int. J. of Mod. Phys. D \textbf{17}, 12, 2255-2259, (2008), [arXiv:0801.1764].
\bibitem{park} M. Park, Phys.~Rev.~D \textbf{78}, 023014 (2008), [arXiv:0804.4331].
\bibitem{simpson} F. Simpson, J. Peacock, A. Heavens, Mon. Not. R. Astro. Soc. \textbf{402}, 3, 2009-2016, (2010), [arXiv:0809.1819].
\bibitem{ak} H. Arakida, M. Kasai, Phys.~Rev.~D \textbf{85}, 023006 (2012), [arXiv:1110.6735].
\bibitem{piattella} O. Piattella, Phys.~Rev.~D \textbf{93}, 129901 (2016), [arXiv:1508.04763v2].
\bibitem{butcher} L. Butcher, (2016), [arXiv:1602.02751].
\bibitem{ll} D. Lebedev and   K. Lake, (2013), [arXiv:1308.4931].
\bibitem{kot} F. Kottler, Ann. Phys. (Leipzig) \textbf{361}, 14, 401–462, (1918).
\bibitem{lake} K. Lake, Phys.~Rev.~D \textbf{65}, 087301 (2002), [arXiv:gr-qc/0103057].
\bibitem{mtw}  C. Misner, K. Thorne, J. Wheeler, \textit{Gravitation} (W. H. Freeman, 1973).
\bibitem{hartle} J. Hartle, \textit{Gravity, An Introduction To Einstein's General Relativity} (Pearson, 2002).
\bibitem{weinberg} S. Weinberg, \textit{Gravitation and Cosmology: Principles and Applications of the General Theory of Relativity} (Wiley, 2013).
\bibitem{wald} R. Wald, \textit{General Relativity} (University Of Chicago Press, 1984).
\bibitem{kkl} V. Kagramanova, J. Kunz, C. Lammerzahl, Phys.~Lett.~B \textbf{634}, 465, (2006).
\bibitem{firstform} K. Pechenick, C. Ftaclas, J. Cohen, Astrophys. J. \textbf{274}, 846 (1983).
\bibitem{defe1} F. de Felice, M. Lattanzi, A. Vecchiato, P. Bernacca, Astron. Astrophys. \textbf{332}, 1133-1141 (1998).
\bibitem{defe2} F. de Felice, P. Bernacca, M. Lattanzi, A. Vecchiato, Astron. Astrophys. \textbf{373}, 336-344 (2001).
\bibitem{oneill} B. O'neill, \textit{Semi-Riemannian Geometry With Applications to Relativity} (Academic Press, 1983).
\bibitem{kriele} M. Kriele, \textit{Spacetime: Foundations of General Relativity and Differential Geometry} (Springer, 2001).
\bibitem{ellis} G. Ellis, \textit{General Relativity and Cosmology}, Proceedings of the International School of Physics, \textbf{47}, 104-182 (Academic Press, New York, 1971).
\bibitem{cohen} C. Ftaclas, J. Cohen, Phys.~Rev.~D \textbf{21}, 8 (1980).
\bibitem{ehlersform} J. Ehlers, Gen.~Rel.~Grav. \textbf{25}, 12, 1225-1266, (1993).
\bibitem{rindler} W. Rindler, \textit{Relativity: Special, General, and Cosmological} (Oxford University Press, New York, 2006).
\bibitem{ehlers} P. Schneider, J. Ehlers, E. Falco, \textit{Gravitational Lenses} (Springer, 2009).
\bibitem{dinverno} R. d'Inverno, \textit{Introducing Einstein's Relativity} (Clarendon Press, 1992).
\bibitem{stephani} H. Stephani, \textit{Relativity: An Introduction to Special and General Relativity} (Cambridge University Press, 2004).
\bibitem{pk} J. Plebanski, A. Krasinski, \textit{An Introduction to General Relativity and Cosmology} (Cambridge University Press, 2006).
\bibitem{irwin} J. Irwin, \textit{Astrophysics: Decoding the Cosmos} (Wiley, 2007).
\bibitem{peacock} J. Peacock, \textit{Cosmological Physics} (Cambridge University Press, 1998).
\bibitem{hendriksen} R. Henriksen, \textit{Practical Relativity: From First Principles to the Theory of Gravity} (Wiley, 2010).
\bibitem{rid} M. Ishak, W. Rindler, J. Dossett, Mon. Not. R. Astro. Soc. \textbf{403}, 4, 2152-2156, (2010), [arXiv:0810.4956].
\bibitem{ri2} M. Ishak, W. Rindler, Gen.~Rel.~Grav. \textbf{42}, 9, 2247-2268, (2010), [arXiv:1006.0014].
\bibitem{ridma} M. Ishak, W. Rindler, J. Dossett, J. Moldenhauer, C. Allison, Mon. Not. R. Astro. Soc. \textbf{388}, 3, 1279-1283, (2008), [arXiv:0710.4726].
\bibitem{ishak1} M. Ishak, Phys.~Rev.~D \textbf{78}, 103006 (2008), [arXiv:0801.3514].
\bibitem{bozza} V. Bozza, Phys.~Rev.~D \textbf{78}, 103005 (2008).
\bibitem{mirag} H. Miraghaei, M. Nouri-Zonoz, Gen.~Rel.~Grav. \textbf{42}, 12, 2947-2956, (2010), [arXiv:0810.2006].
\bibitem{lake2} K. Lake, (2007), [arXiv:0711.0673].
\bibitem{sereno2} M. Sereno, Phys.~Rev.~D \textbf{78}, 083003 (2008), [arXiv:0809.3900].
\bibitem{sereno3} M. Sereno, Phys.~Rev.~Letters \textbf{102}, 021301 (2009), [arXiv:0807.5123].
\bibitem{2} D. Lebedev and K. Lake, Relativistic Aberration and the Cosmological Constant in Gravitational Lensing II. Magnification and time-delay (in preparation).
\end{thebibliography}
\end{document}